On the Reliability of the Geomagnetic Quake as Short Time Earthquake's Precursor for Sofia Region- 2002, 2003

Strachimir Chterev Mavrodiev
mavrodi@inrne.bas.bg
Institute for Nuclear Research and Nuclear Energy, Bulgarian Academy of Sciences, Sofia

**Abstract**
The local "when" earthquake prediction is based on the connection between geomagnetic "quakes" and the next incoming minimum or maximum of tidal gravitational potential. The probability time window for the predicted earthquake is for the tidal minimum approximately +/-1 day and for the maximum- +/-2 days. The preliminary statistic estimation on the basis of distribution of the time difference between occurred and predicted earthquakes for the period 2002- 2003 for Sofia region is given.
The possibility for creating a local "when, where" earthquake research and prediction NETWORK is based on the accurate monitoring of the electromagnetic field with special space and time scales under, on and over the Earth surface. The periodically upgraded information from seismic hazard maps and other standard geodetic information as well as other precursory information is essential.

**Introduction**

The problem of "when, where and how" earthquake prediction can't be solved only on the basis of geodetic data (Aki; Main, 1999a,b; Ludwin, 2001; Pakiser and Shedlock; Geller et al.). The possible tidal trigging of the earthquakes has been investigated for a long period of time (Knopoff, 1964; Ryabl at al., 1968; Shlien, 1972; Molher, 1980, Sounau et al., 1982; Shirley, 1988, Bragin, 1999). The conclusion that the earthquake's time is correlated with the tidal extremums is not unique, because in some of the extremums there are not earthquakes.

The including of additional information in the monitoring, for example, the analysis of the Earth electrical currents signals, permits to estimate the most probable time of incoming earthquake (Thanassoulas 1991, Thanassoulas et al., 2001a, b).

The more accurate space and time measuring set for the Earth's crust condition parameters, the including in the monitoring of the electromagnetic fields measurements under, on and over the Earth surface, the temperature distribution and other possible precursors can be useful for the study of the "when, where and how" earthquake's prediction. For example in the paper Varotsos and Alexopoulos, 1984a, b; Varotsos at al., 1996, the possibility for short-term earthquake prediction in Greece by seismic electric signals was investigated. The results of observations on seismo-electromagnetic waves at two earthquake experimental areas in China were presented in the paper Qian at al., 1994.

The atmospheric and ionospheric electromagnetic phenomena associated with earthquakes were analyzed in many books and papers (Hayakawa and Fujinawa, 1994; Hayakawa et al., 1999, Hayakawa et al., 2000, Hayakawa and Molchanov, 2002) and the future direction of investigation related to earthquake prediction was proposed, as well as its practical application to some concrete events**.** The papers of Oike and Ogawa, 1982, 1994 concern the observations of electromagnetic radiation in the LF and VLF ranges related with the occurrence of earthquake. The results of complex investigation of the variations of crust electrical resistivity as a function of tidal deformations on the basis of extremely low frequency radio station, which permit the hope for increase the reliability of electromagnetic based earthquake prediction, are presented in paper of Saraev et al., 2002.

In papers of Eftaxias 2001, 2002 are presented evidences that electromagnetic anomalyties in wide range of radio frequencies from ULF, VLF to VHF have been observed before some destructive earthquakes in continental Greece.

The impressive results of modified VAN method are presented in papers and web site (ws) (Thanassoulas, 1991; Thanassoulas et al., ws; Thanassoulas et al., 2001 a, b, c, d, e), where, the appropriate measuring of electric Earth signals and their analysis demonstrates that the direction to the epicenter of incoming earthquake can be estimated and the time is defined from the next exstremum of tidal potential. Some possible geophysical models of the phenomena are proposed and the prediction of the future magnitude is analyzed. The including of more than one site in the monitoring will permit short term earthquake prediction and will give some estimation for the magnitude (Thanassoulas, ws).

In site Ustundag and papers cited there, are presented the results of electropotential monitoring, based on the special constructed electrometer and appropriate temporal data acquisition system, for researching the electropotential variations as earthquake precursor.
One have to mention the satellite possibilities for monitoring the radiation activity of the Earth surface for discovering the anomalyties, which should be the earthquake precursor: Dean, ws.

The analyses of the data from satellite monitoring for the ionosphere and the Earth radiation belt parameters also give evidences for anomalyties which can be interpreted as earthquake precursor. The information for the last results from the developing of earthquake precursor research could be found in the conference sites: Contadakis, 2002 and Papadopoulos, 2003.
The data for the connection between incoming earthquake and meteorology effects like quasistationary earthquakes clouds can be seen in the site Zhonghao. The statistic from 1993 for the reliability of prediction is also represented together with some theoretical models and estimations for the effect.

In order to summarize the results we can say, that the standard geodetic monitoring USGS Pf ws, the monitoring of different component of electromagnetic field under, on and over the Earth surface, some of the atmospheric anomalyties and behavior of charge distribution in the Earth radiation belts (see for example Silina, 2001; Larkina and Ruzhin, 2003), sometime could serve as unique earthquake's precursor. It is obvious, that for solving the reliability problem different approaches should be unified, including the biological precursor data.

The progress in geomagnetic quake earthquake precursor approach (Mavrodiev, 2001) is presented (Mavrodiev, 2002 a, b, 2003 a, b). The approach is based on the understanding that earthquake processes have a complex origin. Without creating of adequate physical model of the Earth existence, the gravitational and electromagnetic interactions, which ensure the stability of the Sun system and its planets for a long time, the Earth prediction problem can not be solved. The earthquake part of the model can be repeated in the infinity way "theory- experiment- theory" using nonlinear inverse problem methods looking for the correlations between fields in dynamically changed space and time scales. Of course, every approximate model (see for example Thanassoulas, 1991; Thanassoulas et al., 2001a, b) which has some experimental evidence has to be included in the analysis. It seems obvious that the problem of adequate physical understanding of the correlations between electromagnetic precursors, tidal extremums and incoming earthquake is connected with the progress of the Earth magnetism theory.

The achievement of the Earth surface tidal potential modelling, which includes the ocean and atmosphere tidal influences, is an essential part of the research. In this sense the comparison of the Earth tides analysis programs (Dierks and Neumeyer, ws) for the ANALYZE from the ETERNA-package, version 3.30 (Wenzel, 1996a, b), program BAYTAP-G in the version from 15.11.1999 (Tamura, 1991), Program VAV (version from April 2002) - Venedikov et al, 2001, 2002 is very useful.

The role of geomagnetic variations as precursor could be explained by the obvious hypothesis that during the time before the earthquakes, with the strain, deformation or displacement grows in the crust in some interval of density changing, arise the chemical phase shift which leads to an electrical charge shift. The preliminary Fourier analysis of geomagnetic field gives the time period of alteration in minute scale. Such specific geomagnetic variation we call geomagnetic quake. The piezo- effect model for electrical currents can not explain the alternations due to its linearity.

K-index (Balsh), accepted for the geomagnetic conditions, can not indicate well the local geomagnetic variation for time minutes period, because it is calculated on the basis of 3 hours data. Nevertheless, the K- index behavior in the near space has to be analyzed because of the possible Sun wind influence on the local behavior of geomagnetic filed. If the field components are measured many times per second, one can calculate the frequency dependence of full geomagnetic intensity and to analyze the frequency spectrum of geomagnetic quake. If the variations are bigger than usual for some period of time, one can say that we have the geomagnetic quake, which is the earthquake precursor. The nonlinear inverse problem analysis for 1999- 2001 of geomagnetic and earthquake data for Sofia region gives an estimation, that the probability time window for the predicted earthquake (event, events) is approximately minimum +/-1 day and maximum +/-2 days of the Earth for the tidal potential .

The future epicenter coordinates could be estimated from the data from at least 3 points of measuring the geomagnetic vector, using the inverse problem methods, applied for the estimation the coordinates of the volume, where the phase shift arrived in the framework of its time window.

In the case of incoming big earthquake (magnitude > 5 - 6) the changes of vertical electropotential distribution, the Earth temperature, the infrared Earth radiation, the behavior of water sources, its chemistry and radioactivity, the atmosphere conditions (earthquakes clouds, etc.) and the charge density of the Earth radiation belt, have to be dramatically changed near the epicenter area.

The achievements of tidal potential modeling of the Earth surface, including ocean and atmosphere tidal influences, multi- component correlation analysis and nonlinear inverse problem methods in fluids dynamics and electrodynamics are crucial for every single step of the constructing of the mathematical and physical models.

In Part 1 are given the 2002- 2003 statistics estimations for the reliability of the time window earthquake prediction on the basis of geomagnetic field measurements (Mavrodiev, 2002b) and the Earth tidal behavior Venedikov at al., 2002] for Sofia region

In Part 2 the posterior analysis for Alaska, 2002, Magnitude 8.2 earthquake on the basis of seconds CMO geomagnetic data and for Hokkaido 2003, MMB minute geomagnetic data is presented.

In Part 3 a short proposal for creating of Short Time Earthquake prediction in local NETWORK is given.

**Part 1. The geomagnetic field quake as a time window earthquake's precursor for Balkan, Black Sea region**

It is useful to stress that the author's interests to the earthquake's prediction problem arise as a result of complex research of the Black Sea ecosystem about 15 years ago (Mavrodiev, 1998). During the time of gathering the historical data for the ecosystem was observed, that the Crime earthquake in 1928 occurred, as an evidence for electromagnetic and earthquake correlations. Such hypothesis has been proposed by the academician Popov in the early 20- 30-ties of the last century (private communication).

According the INTERMAGNET requirements for measuring the geomagnetic field (see Figure 1) on Earth surface (http://www.intermagnet.org), the accuracy is ±10 nT for 95% of reported data and ±5 nT for definitive data, with one sample per 5 seconds, in the case of Vector magnetometer (**F** (XYZ) or **F** (HDZ)) and 1 nT, with 3 samples per second, for Scalar Magnetometer (F)**.**

The geomagnetic vector projection H is measured with relative accuracy less or equal to1 nT by a fluxgate, feedback based device of rather original and simple, but powerful construction. (know- how of JINR, Dubna, Boris Vasiliev, 1998, private communication). It is used with 2.4 samples per second. Due to technical reasons the sensor was oriented under the Horizon in a manner that the measured value of H is around 20000 nT (see Figures 1 and 2).

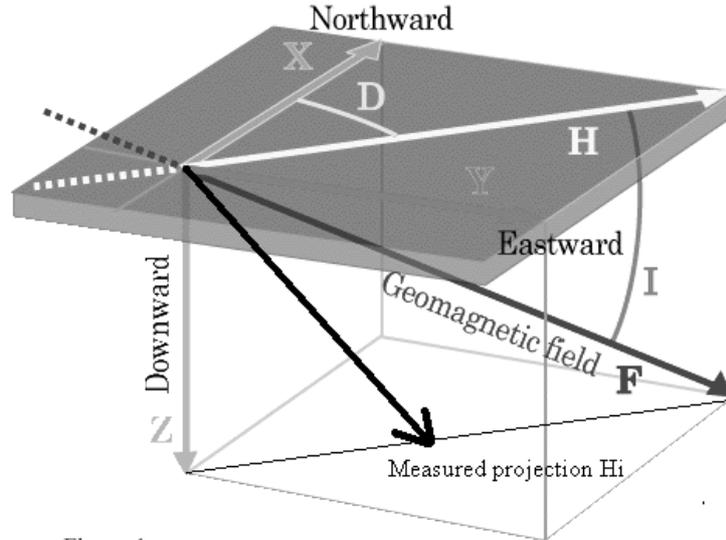

Figure 1
The components of Earth Geomagnetic field
Source: http://swdcwww.kugi.kyoto-u.ac.jp/element/eleexp.html

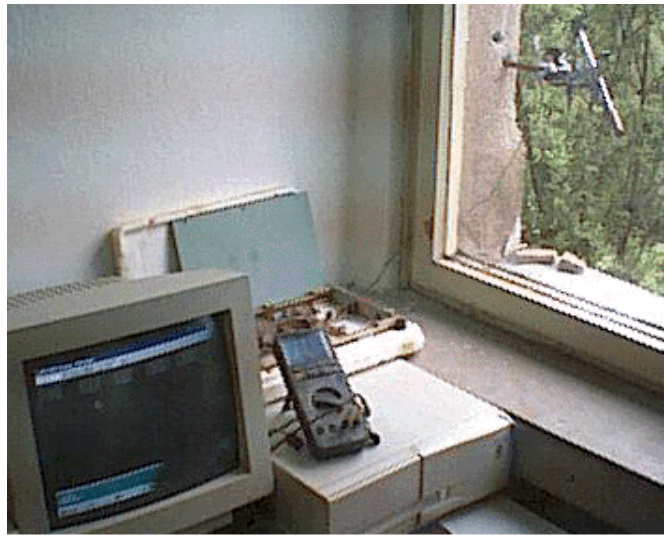

Figure 2 The Magnetometer with Sensor and PC

The minute averaged value $H_m$ and its error $\Delta \overline{H}_m$ are

$$\overline{H}_m = \sum_{i=1}^{N_m} \frac{H_i}{N_m} \text{ and } \Delta \overline{H}_m = \sum_{i=1}^{N_m} \frac{\Delta H_i}{N_m},$$

where $H_i$, $\Delta H_i$ are the measured Nm = 144 times per values of the field and their experimental error.

The standard deviation $\sigma_{H_m}$ and its error $\Delta \sigma_{H_m}$ for every minute are

$$\sigma_{H_m} = \frac{\sqrt{\sum_{i=1}^{N_m}(H_i - \overline{H_m})^2}}{N_m} \text{ and } \Delta \sigma_{H_m} = \frac{\sqrt{\sum_{i=1}^{N_m}(\Delta H_i - \overline{\Delta H_m})^2}}{N_m}$$

Looking for the correlations between the behavior of the geomagnetic field, Earth tidal gravitational potential and the occurred earthquakes (from 1999 to 2001) it turns out that the daily averaged value of $\sigma_{H_m}$ and $\sigma_{\Delta H_m}$, which we denote by Sig (ΔSig), is playing the role of earthquake precursor.

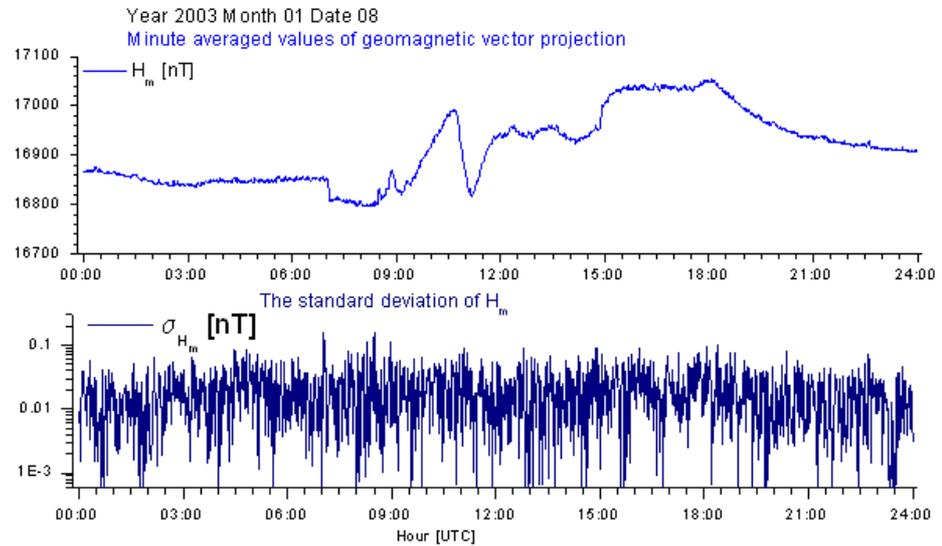

Figure 3
The behavior of geomagnetic field for a normal day

Figure 3 illustrates the behavior of geomagnetic field component $H_m$ and its variation $\sigma_{H_m}$ for a period without earthquakes precursor in the region. Figures 4 and 5 illustrate the behavior of geomagnetic field and its variation, which is unusual. Is this case there is a geomagnetic quake, which is precursor for incoming event (earthquake or earthquakes). One has to be sure that there are not a cosmos or Sun wind reasons for the geomagnetic quake (see the sites [NOAA ws].

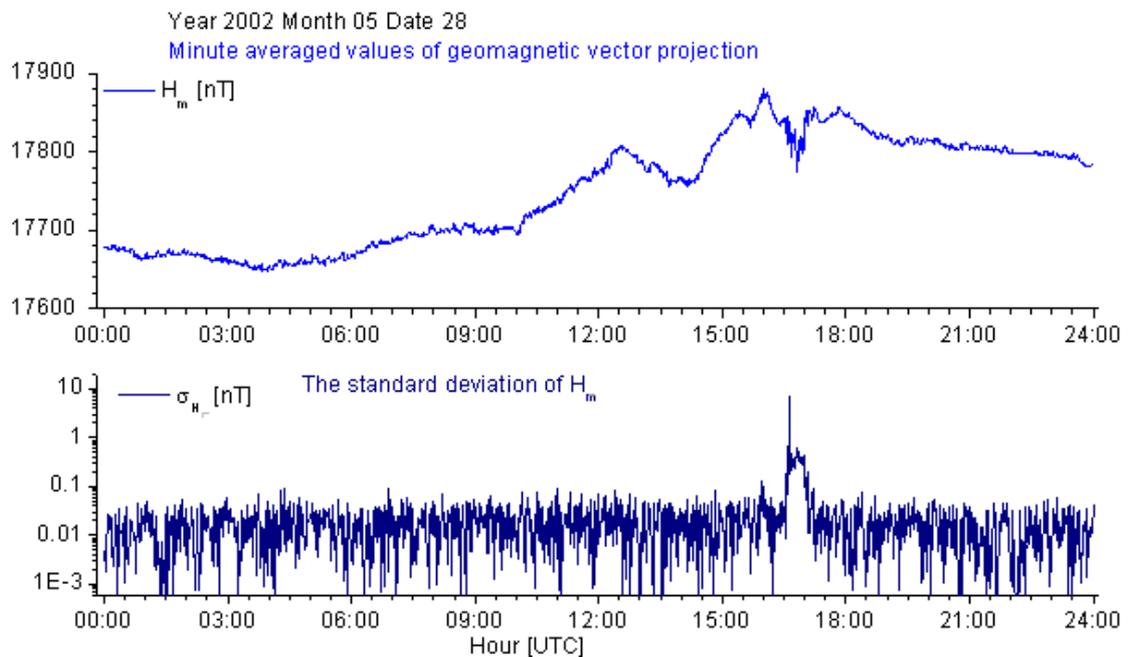

Figure 4
The behavior of geomagnetic field for a day with a signal for a near future earthquake

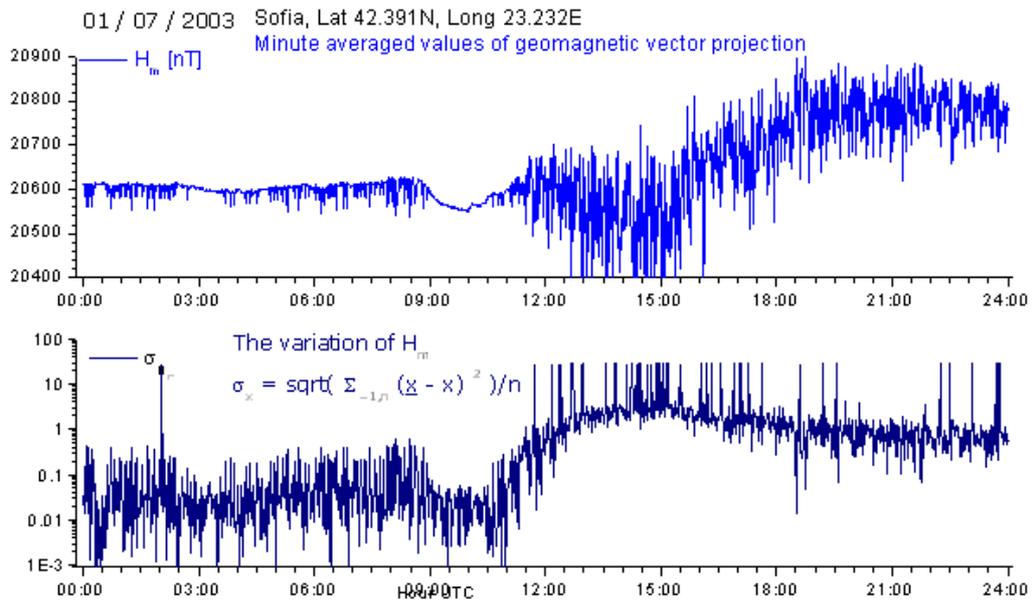

Figure 5
A day with geomagnetic quake, which is eq's precursor

For example in Figure 4 the predicted time was 3 +/- 1 June, 2002 and the prediction was confirmed with earthquake, occurred at 03/06/2002 02:04, Lat41.95N, Lon23.10E, Dep8, Mag2.6, Ml, 50 km from Sofia, $S_{ChtM}$ = 598 [Mag/r$^2$].

The preliminary Fourier analysis of $H_m$ data gives the fact that the bigger geomagnetic variations are cased from the arriving for hours time period new frequencies with periods from 10-th of seconds till 10-th of minutes and with very specific amplitude behavior. Such spectrum, which arrives for hour period of time, is invisible for minute samples measuring.

The probability time window of the incoming event (or events) is defined by the next date of the Earth tidal potential extremum with tolerance approximately for the tidal minimum +/-1 day and for the maximum +/-2 days- Figure 6.

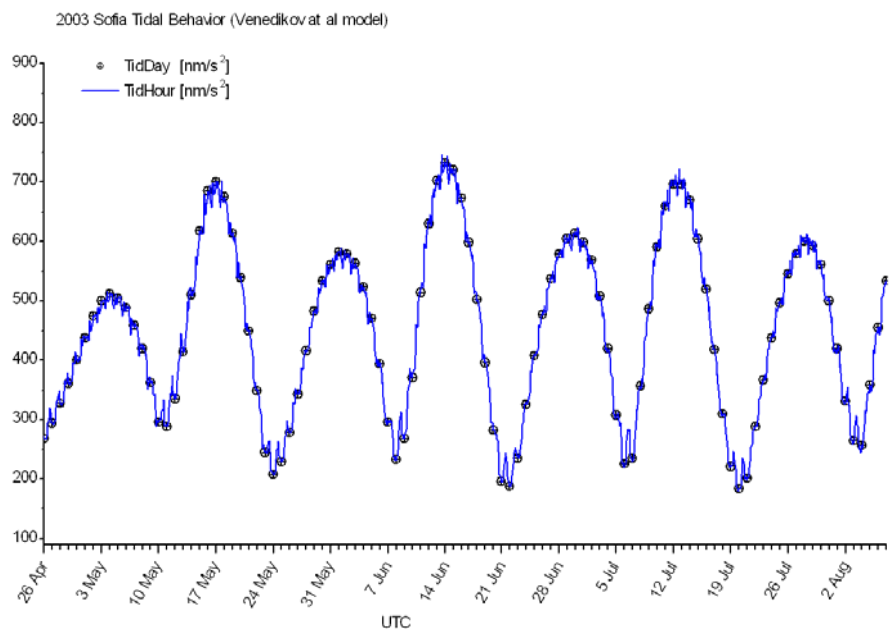

Figure 6
The Tidal Gravitational Potencial behavior for June, 2003
Vinedikov at al model[ ]

The uncertainty of distinguishing the predicted event (or group of events – for example aftershocks) from the events which occurred in the region at different distances and magnitudes in the predicted time window is solved on the basis of inverse problem methods with new earthquake's influence characteristic function $S_{ChtM}$:

$$S_{ChtM} = \frac{2 Magnitude}{(R_{eq} + Distance)^2}, R_{eq} = 0.040 + \frac{Dept}{Magnitude} \text{ [Thousand km]}$$

The physical sense of the function $S_{ChtM}$ is a density distribution on the Earth surface of earthquake's magnitude [magnitude vs. distribution]. In the point of measurement $S_{ChtM}$ is logarithmically proportional to the energy influence of the earthquakes. It is important to point out that the first consideration of the magnitude and distance dependences was obtained on the basis of nonlinear inverse problem methods. Obviously, the nearer and biggest earthquake (relatively biggest value of $S_{ChtM}$) will bear more electropotential variations, which will generate more power geomagnetic quake. At this stage of the study, as a measure of daily geomagnetic state can serve the value of averaged for 24 hours (1440 minutes) standard deviation $\sigma_{Hm}$:

$$(1) \quad Sig = \sum_{i=1}^{1440} \frac{\sigma_{H_m}}{1440}, \quad \Delta Sig = \sum_{i=1}^{1440} \frac{\Delta \sigma_{H_m}}{1440}$$

**The simple and usually working criterion for evidence of geomagnetic quake is when Sig increases for two consecutive days i, i+1 and the differences between the values of Sig are bigger than the mean arithmetic sum of their errors ΔSig:**

**(2)      Abs ($Sig_{i+1} - Sig_i$) > ($\Delta Sig_i + \Delta Sig_{i+1}$)/2.**

**If criteria (2) are fulfilled and there are not a Cosmos and Sun generated variations of geomagnetic field, it could be concluded that the geomagnetic quake has happened. Such quake is unique precursor for incoming earthquake and in the next minimum or maximum of the local Tidal gravitational potential somewhere in the region this predicted earthquake will occur.**

**For some of the cases, the criteria (2) have to be calculated for decades of hours or ten's of minutes.**

**The signal for earthquakes with different epicenters we observe in the case when the specific behavior of field and its standard deviation occurs more than ones per day at different hours.**

**The analysis of the precursor function Sig on the basis of special digital 5 points derivatives can serve in the future for creating the algorithm for automated alert system.**

It is obvious that the more detailed time window can be achieved by analyzing the daily variations of tidal potential, calculated every hour.

As an example, the parameters of predicted events (Figure 5) are presented in Table 1.

**Table 1**

| DDMMYYhhmm | Latitude | Longitude | Dept[km] | Magnitude | Distance [hundred km] | $S_{ChtM}$ |
|---|---|---|---|---|---|---|
| 03/07/2003 15:49 | 42.96 | 25.29 | 10 | 2.6 | 1.10 | 219 |
| 03/07/2003 20:51 | 41.96 | 23.28 | 12 | 2.9 | 0.48 | 684 |
| 05/07/2003 21:58 | 40.35 | 26.14 | 2 | 4.0 | 2.97 | 70 |
| 06/07/2003 19:10 | 40.46 | 26.01 | 10 | 5.7 | 3.16 | 89 |
| 06/07/2003 19:39 | 40.62 | 25.25 | 2 | 4.2 | 2.35 | 111 |
| 06/07/2003 20:02 | 40.72 | 25.98 | 20 | 3.2 | 2.54 | 71 |
| 06/07/2003 20:10 | 40.46 | 26.08 | 10 | 5.0 | 3.20 | 76 |
| 06/07/2003 20:48 | 40.28 | 26.08 | 2 | 4.1 | 3.01 | 70 |
| 06/07/2003 21:58 | 40.38 | 26.10 | 10 | 3.9 | 2.92 | 70 |
| 06/07/2003 22:05 | 40.34 | 26.00 | 2 | 3.9 | 2.92 | 71 |
| 06/07/2003 22:42 | 40.95 | 26.00 | 10 | 4.6 | 2.80 | 89 |
| 07/07/2003 00:24 | 40.25 | 25.98 | 2 | 3.7 | 3.00 | 64 |
| 07/07/2003 00:48 | 40.44 | 25.87 | 18 | 3.4 | 2.77 | 66 |
| 07/07/2003 07:15 | 41.67 | 24.88 | 10 | 3.1 | 1.21 | 230 |
| 07/07/2003 16:17 | 40.37 | 25.91 | 10 | 3.3 | 2.85 | 61 |
| 08/07/2003 02:48 | 41.82 | 22.93 | 12 | 2.8 | 0.65 | 465 |
| 08/07/2003 12:00 | 42.84 | 23.32 | 10 | 2.5 | 0.50 | 565 |

At this stage of the study all earthquakes have the same $S_{ChtM}$ for different definitions of the magnitude. After the developing of mathematical models of empirical and theoretical dependences between incoming earthquake processes, magnetic quake and parameters of earthquake on the basis of inverse nonlinear problem we will obtain a set of $S_{ChtM}$ functions in correspondence with the different definition of magnitude. The volumes, its depth, chemical and geological structures of the region have to be included in the dependences as well.

Figure 7 illustrates the possibility for registration of big world earthquakes by specific behavior of the geomagnetic change.

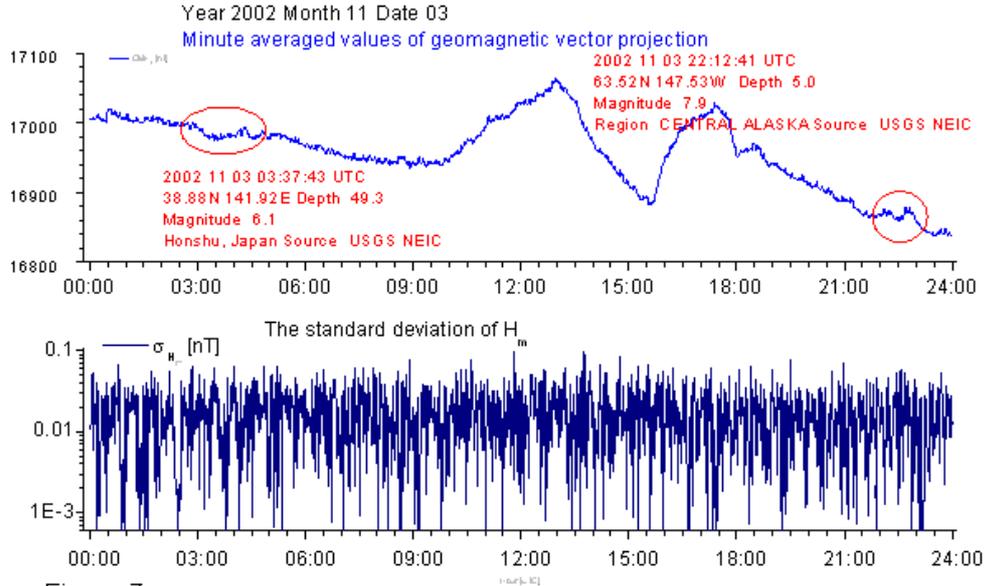

Figure 7
A registration of big world earthquake, Mag>6

In Figures 8, 9, 10 and 11 the tidal potential, $S_{ChtM}$, magnitude, distance from Sofia and function Sig are presented for every quarter of 2003. In the text boxes are the data of the geomagnetic quake (date of the precursory signal) and the time window for predicted event (events).

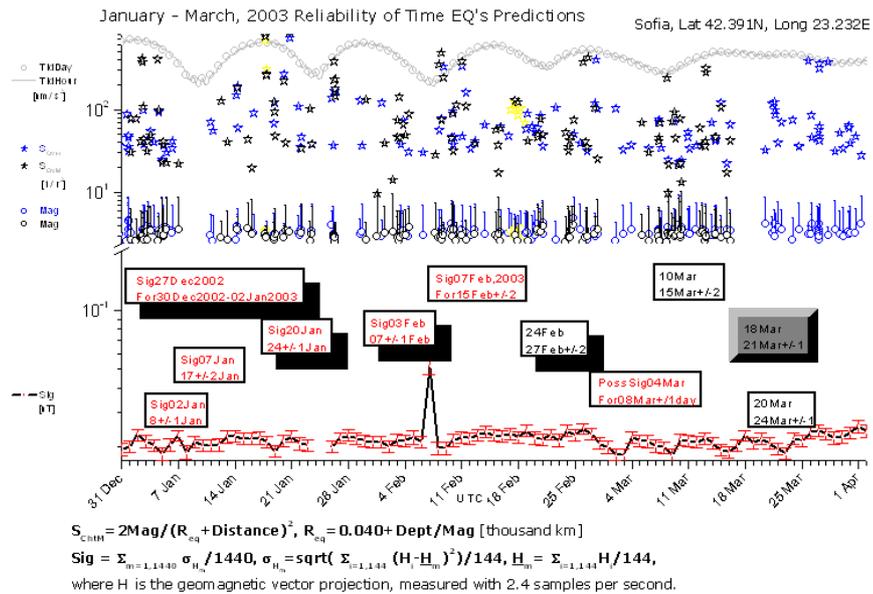

Figure 8

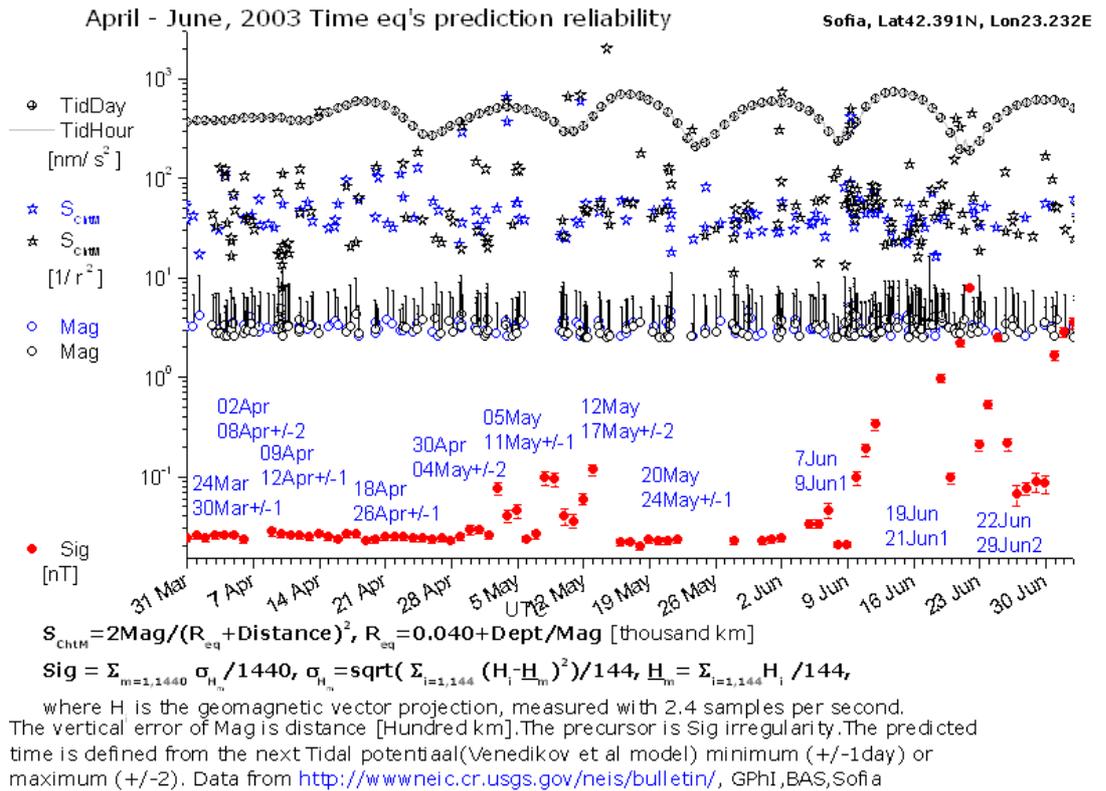

Figure 9

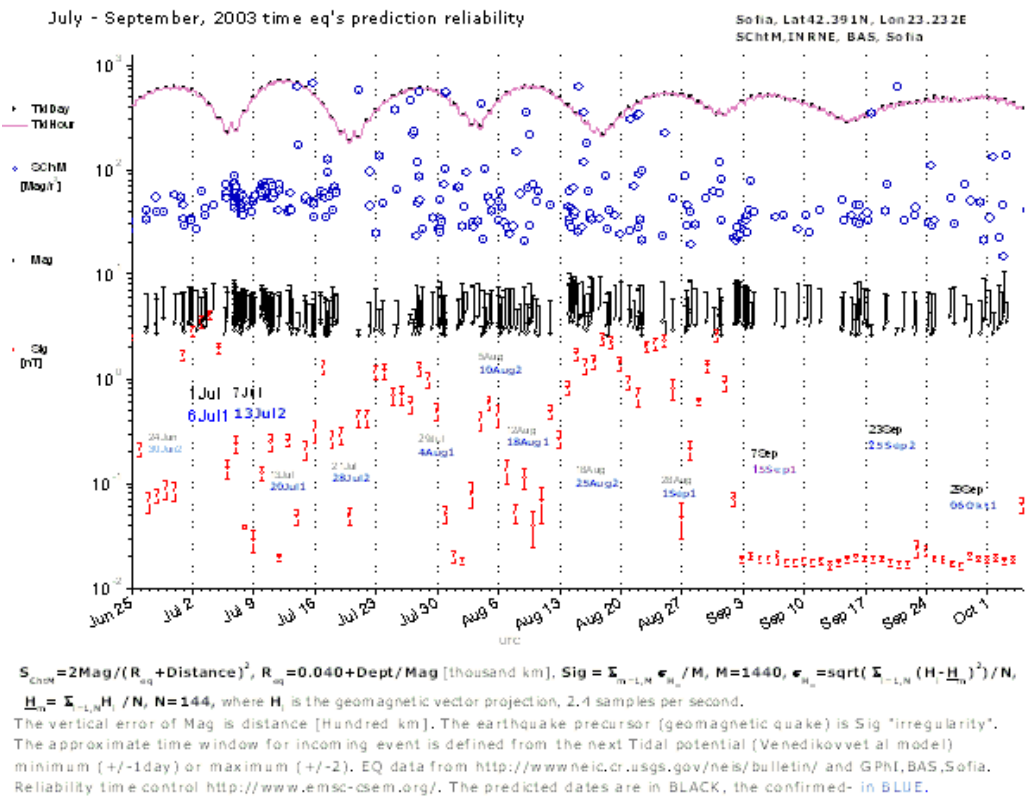

Figure 10

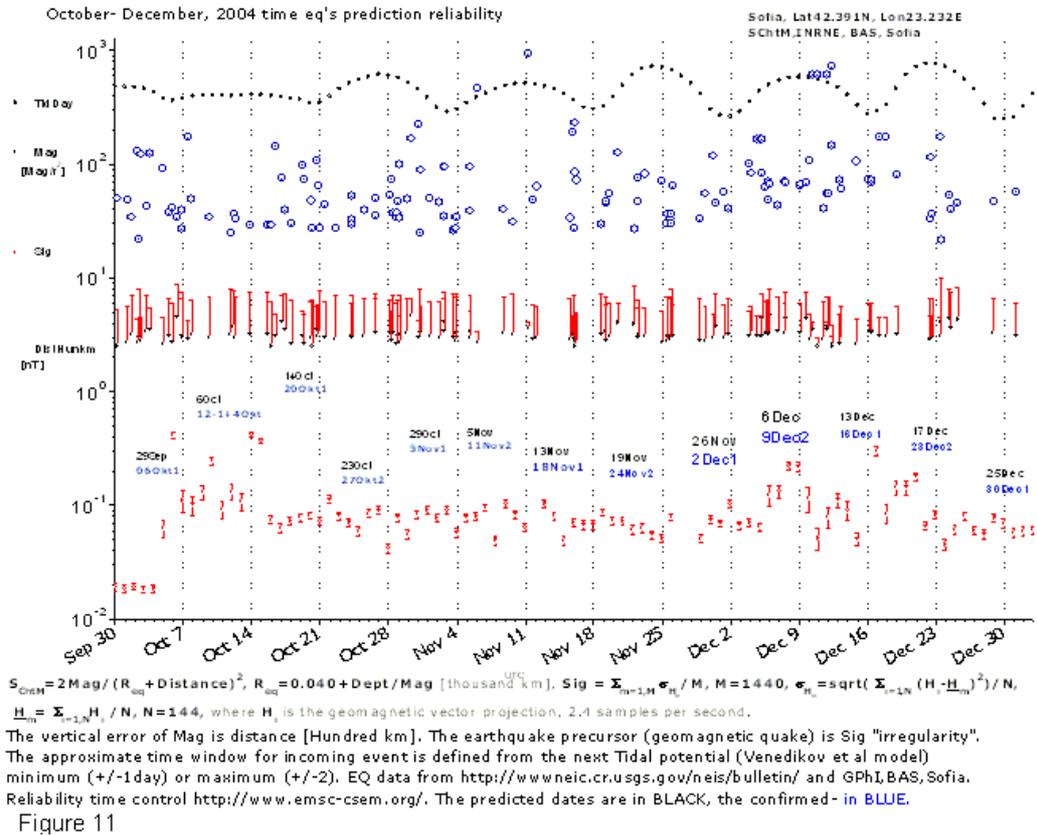

Figure 11

In Figure 12 the distributions of the difference between the times of predicted events, which occurred, calculated for 6, 12, 18 and 24 months (starting from January, 2002), are presented. The distribution growth, without widening and its approximation to the Gaussian distribution with the time, is an argument for the causality- consequences origin between the correlation geomagnetic signal- tidal potential exstremum and occurred earthquake. The number of earthquakes in Figure 12 is greater than the number of the predictions for the events. The explanation of this is that some earthquakes with greater magnitude are following by aftershocks.

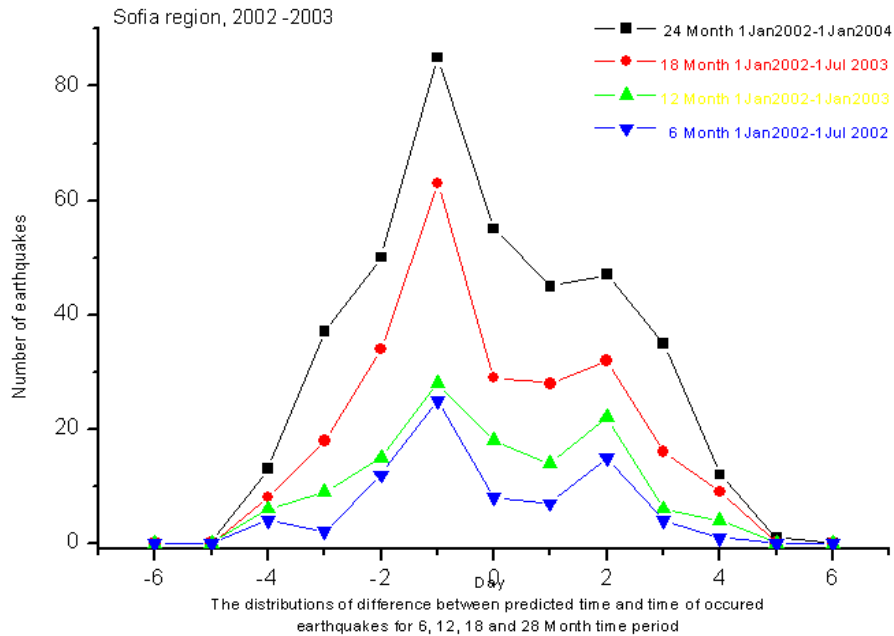

Figure 12

In Figure 13 the distributions of the difference between times of all occurred earthquakes in the region (distance less than 500 km), magnitude grater then 2.6 and the corresponding tidal extremums time, calculated for 6, 12, 18 and 24 months (starting from January, 2002) as well the comparison with Figure 12 are presented. The fact, that the distributions of all earthquakes are flatter than the distributions of predicted earthquakes, can be consider as evidence that the correlation between geomagnetic quake and tidal extremum in the framework of physical mean of function $S_{ChtM}$ is a reliable precursor.

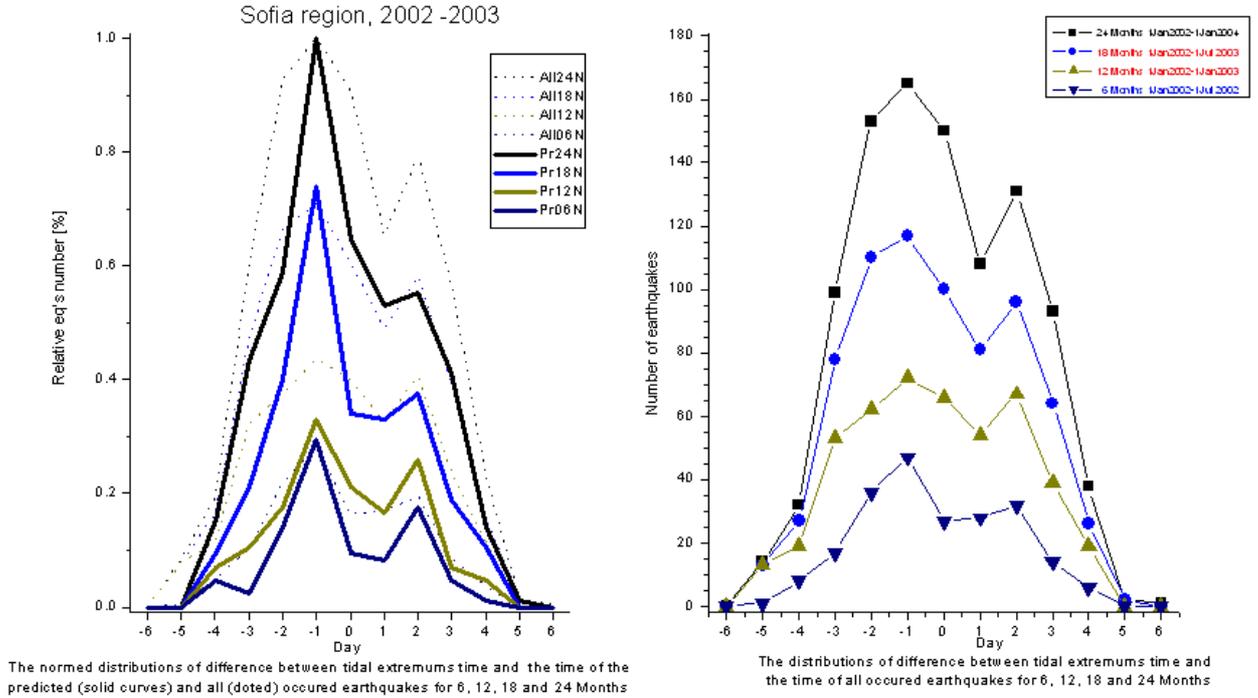

Figure 13. The normed distributions of difference between tidal extremums time and the time of the predicted (solid curves) and all (doted) occured earthquakes for 6, 12, 18 and 24 Months. The distributions of difference between tidal extremums time and the time of all occured earthquakes for 6, 12, 18 and 24 Months.

Next Figure 14 expresses the obvious fact that the incoming earthquake with grater magnitude can be predicted at greater distances. Nowadays estimation is that bigger earthquakes (magnitude >5) could be predicted for distances till 500- 600 km.

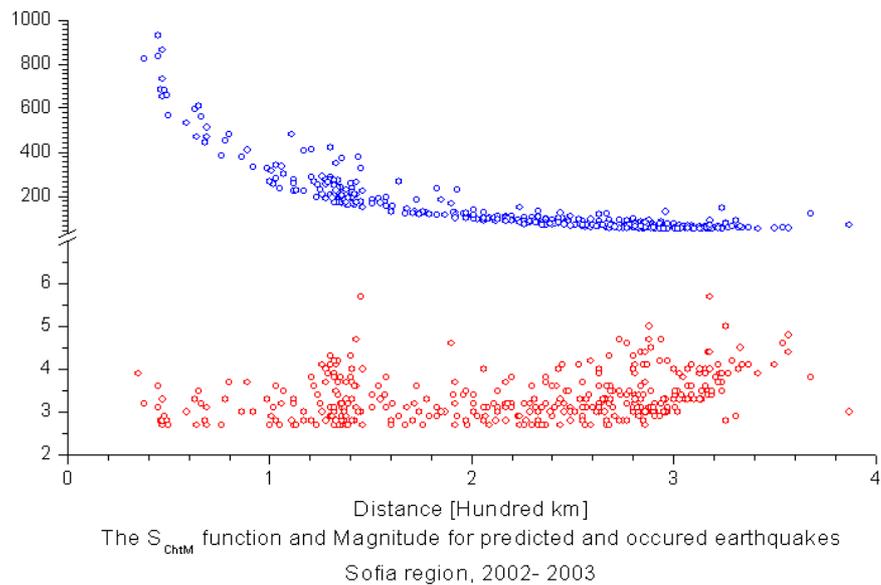

Figure 14. The $S_{ChtM}$ function and Magnitude for predicted and occured earthquakes. Sofia region, 2002- 2003. $S_{ChtM} = Magnitude/(R_{eq}+Distance)^2$, $R_{eq}=0.040+Magnitude/Dept$ [Thousand km]

In Figure 15 is represented the magnitude distribution of the predicted events- 380. It could be seen that the earthquakes with magnitude less then 3 are a small part (26 %) of all predicted and occurred earthquakes.

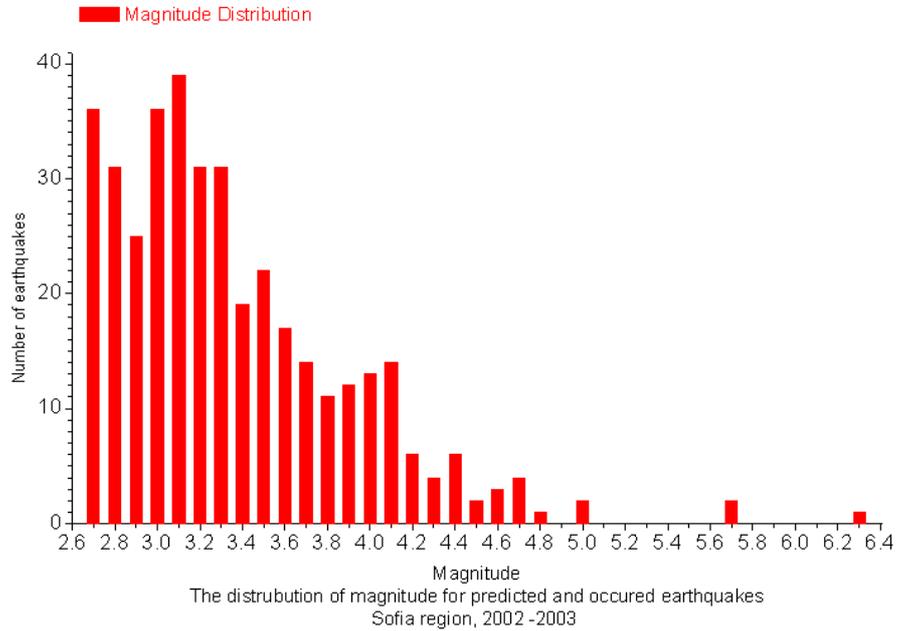

Figure 15

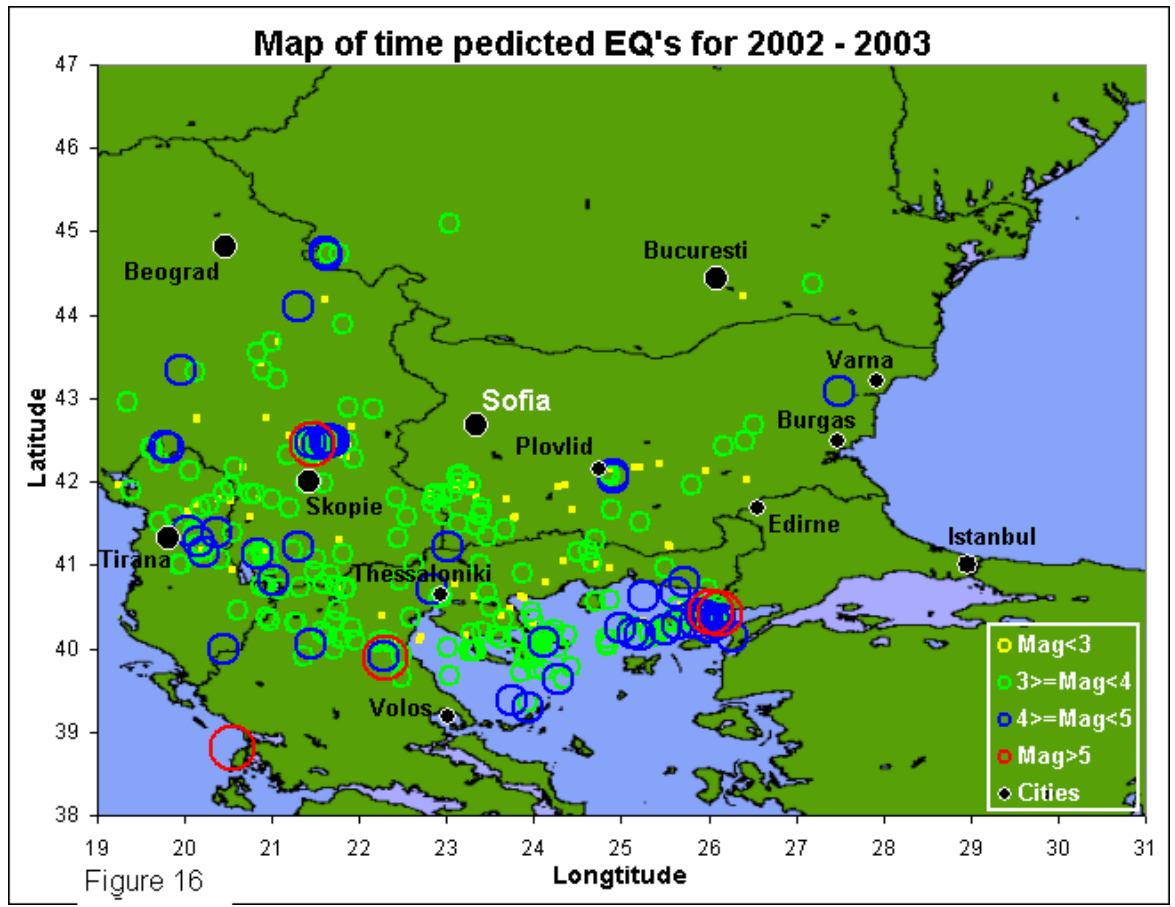

Figure 16

The independent control (in the framework of Strasbourg recommendations about earthquake prediction of the European Union for ethical and public security reasons) of the time window earthquake prediction reliability was organized in the framework of the Bulgarian Academy of Sciences, its Geophysical Institute and colleagues from Greece and Turkey, which are interested in this topic of research (Tsatsaragos, ws; Ustundag, ws), starting from January 2002.

As a conclusion, from the observations for the analyzed period 2002- 2003 there were 110 extremums of tidal potential and 96 events (486 for 2002 and 552 for 2003). For 85 of them the time window was successfully predicted. For two of occurred events the precursor quake was not established (in the first month of the reported time period) by the everyday analysis and 9 events were not predicted, because of hardware problem. The total number of earthquakes with magnitude grater then 2.5 was 1038. The epicenter of the incoming event can not be estimated using one point geomagnetic data, so one can not number the really predicted earthquakes. But there are visible correlations between hazard risk estimation map (Giardini et al., 2003) and Figure 16.

The above results for reliability time window earthquake prediction can be consider as a first preliminary step for solution of "when, where and how" earthquake prediction at level "when", using the geomagnetic measurements.

The first prove, that in the framework of such complex approach, the "when, where and how" earthquake prediction problem can be solve will be the "when, where" prediction on the basis of at least 3 points for electromagnetic real time monitoring. If the statistic estimation is successful for a long enough period of time (6- 12 months) and the established correlations are confirmed by the adequate physical model solutions, one could say that the earthquake prediction problem is under solving using the geomagnetic quake precursor.

**Part 2. The posteriori analysis**

**Alaska, 2002 Magnitude 8.2 earthquake, Seconds CMO geomagnetic data**

In the case of geomagnetic vector measurements the precursor signal Sig is defined as daily averaged sum of normalized standard deviations $\dfrac{\sigma_{\overline{H_m}}}{\overline{H_m}}$, $\dfrac{\sigma_{\overline{D_m}}}{\overline{D_m}}$, $\dfrac{\sigma_{\overline{Z_m}}}{\overline{Z_m}}$ (Figure 17).

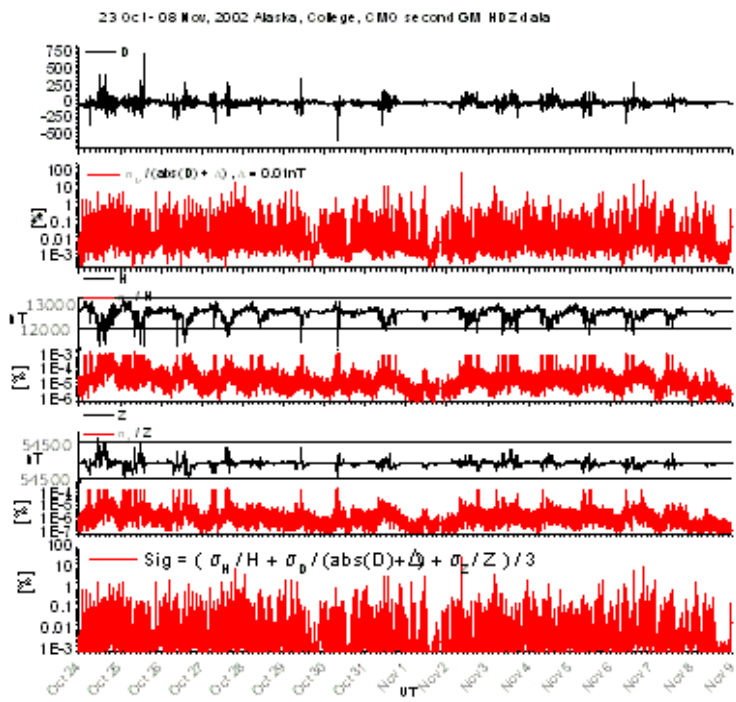

The minute avereged second HDZ data and their normed standard deviation
Figure 17

The data for tidal daily behavior, function $S_{ChtM}$, the magnitude of occurred earthquakes and precursor signal **Sig** are presented in Figure 18 for the period from 24 October to 9 November, 2002. The precursor **Sig** is calculated using the second HDZ College Geomagnetic observatory data (CMO Intermagnet geomagnetic observatory with coordinates 64.84N, 148.86W.

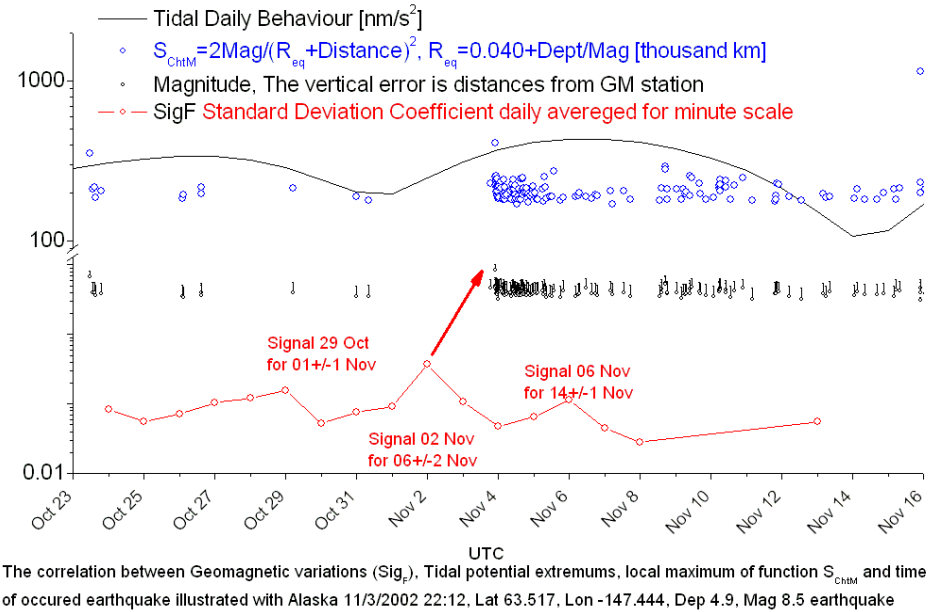

Figure 18

Although the geomagnetic data are only seconds, the correlation between geomagnetic precursor function and incoming earthquake is clearly seen for 3 events, including the big earthquake with magnitude = 8.5. The second CMO geomagnetic data were kindly given by USA Intermagnet group.

**Hokkaido 2003, MMB minute geomagnetic data**

The preliminary analysis on the basis of Sofia region experience of the second MMB Intermagnet data shows that the Hokkaido could be a good polygon for testing the possibility for ``when, where'' eq's prediction if two more Intermagnet geomagnetic sites F(HDZ) with at least 10 samples per second will be created.

Analogous analysis for the regions, where there are Intermagnet geomagnetic stations, for establishing the correlation between local geomagnetic quake and incoming, in time window, defined from the time of next minimum or maximum of the Tidal behavior, earthquakes, was performed for England, India, Turkey regions (Mavrodiev 2003 a, b).

**Part 3. Proposal for creating of Short Time Earthquake prediction local NETWORK**

We will not discuss the long time prognostic system for estimation of earthquake risk. According to Keilis-Borok, ws and Giardini et al., 2003 they are well known.

The aim of this Proposal is to create a system for study of the reliability of the local forecast system for earthquakes in the interval Mag > 2.6- 3 and radius till 600 km. The system is complex and the attended practical result will be a short time "when, where" eq's prediction. The problem "when, where and how" will be solved step by step in creating an adequate physical theoretical model for the Earth magnetism. For such complex research a new type of scientific unification has to be realized including experimental, theoretical and technological parts:

**Experimental data for:**

Geomagnetic field, Electro-potential distribution in the Earth crust and atmosphere, Temperature Earth crust distribution, Crust parameters (strain, deformation, displacement), Gravitational anomaly map, Season and day independent depth temperature distribution, Water sources parameters (debit, temperature, chemical composition, radioactivity), Gas emissions, Ionosphere condition parameters, Infrared radiation of Earth surface, earthquake clouds, Earth radiation belt, Sun wind, Biological precursors.

**Theory**: The achievements of tidal potential modeling of Earth surface including ocean and atmosphere tidal influences, multi- component correlation analyses, nonlinear inverse problem methods in fluids dynamics and Maxwell equations are crucial.

**Technologies: GIS** for archiving, analysis, visualization and interpretation of the data and non-linear inverse problem methods for creating of theoretical models for the parameters behavior, correlations and dynamics.

The set of the devices used, has to be in correspondence with known data for earthquakes risk zone (gravitational anomalyties, crust parameters monitoring-strain, deformation, displacement and seismic hazard evaluation maps). The geomagnetic device set distance has to be 150 – 200 km, the electro- potential from 100 to 200 km in dependence of present geological situation and its history. The set for monitoring of the daily and season crust temperatures has to be in the range of 100 km. The correlations with Sun wind influence have to be estimated in real time.

The system has to be created step by step. The condition for next step has to be build on physically clear new theoretical correlations or dynamical models and, of course, the successful "when", "when, where" or "when, where and how" earthquake predictions.

Finally, the number of world earthquake's with magnitude greater then 4 are represented in Fig. 19. The correlation between the eq's number, global warming and consequent increasing the sea level and the amplitude of ocean tides is obvious. The possible explanation is in the fact, that the global warming, because of ice decreasing, is a reason for perturbation of the momentum equilibrium of the Earth and of the Ocean tidal amplitudes increasing. So, the continents movements are activated in such manner, that the number of earthquakes increase. **The fact, that the number of earthquakes with magnitude grater then 7, conserve its chaotic character, can be interpreted as a chance that we have still some time to stop the global warming and to prevent incoming anthropogenic geodetic disasters.**

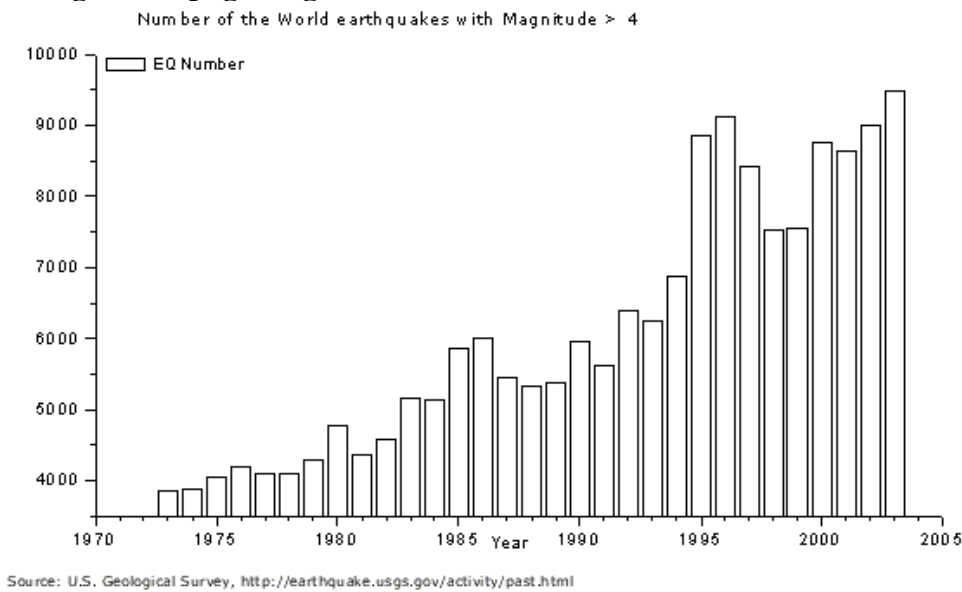

The number of earthquake's with Magnitude > 4
Figure 19

### Conclusion

The correlations between the local geomagnetic quake and incoming earthquakes, which occur in the time window defined from tidal minimum (+/- 1 day) or maximum (+/-2 days) of the Earth tidal gravitational potential are tested statistically. The distribution of the time difference between predicted and occurred events is going to be Gaussian with the increasing of the statistics.

This result can be interpreted as a possible first reliable approach for solving the "when" earthquakes prediction problem using the geomagnetic data.

On the basis of electromagnetic monitoring under, on and over Earth surface a research for solution of "when, where" earthquake prediction problem is proposed. Under the hypothesis that the current, which born the geomagnetic quake, has bigger vertical component, the data of two geomagnetic vector devices are enough for determination of the future epicenter. Three devices equipment will permit to study the correlation between Earth surface distribution of precursor function Sig and the magnitude of the incoming earthquake.

In the last Figure 20 is presented the reliability test for time prediction, using the new $S_{ChtM}$ function [Energy / squared distance] for 2003.

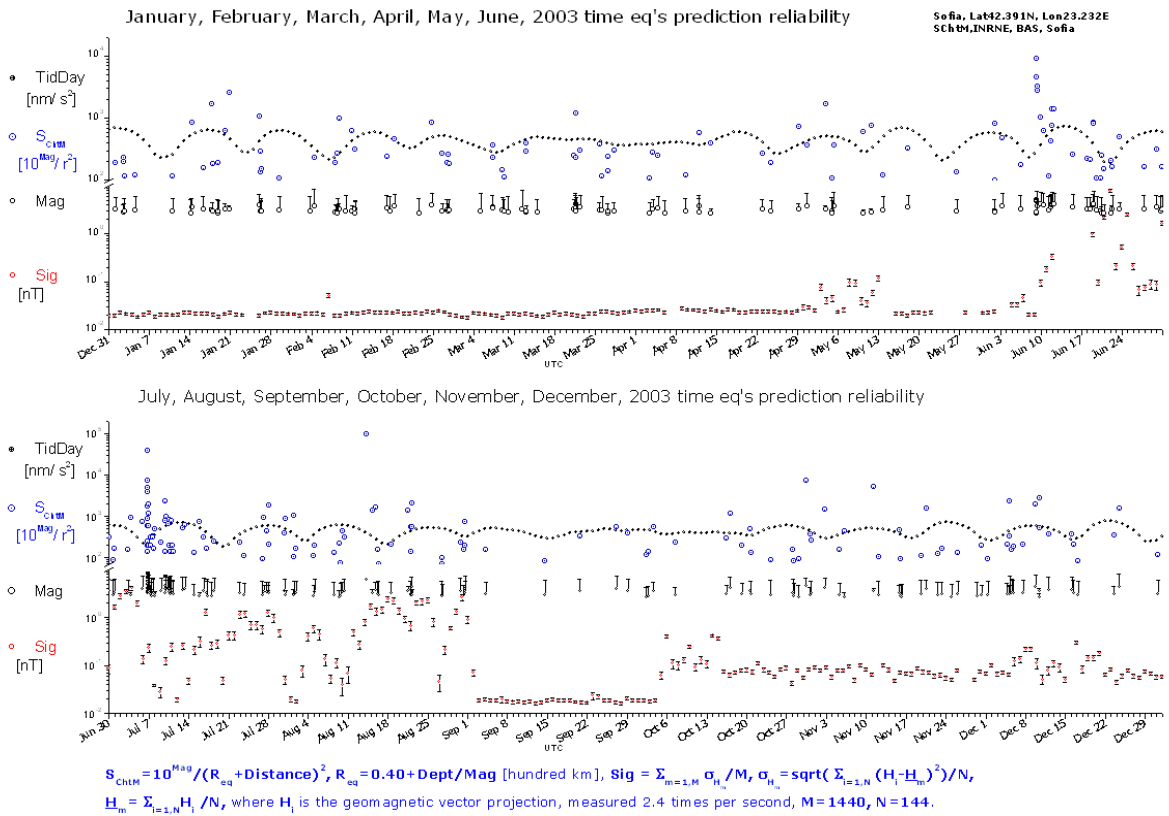

$S_{ChtM} = 10^{Mag}/(R_{eq}+Distance)^2$, $R_{eq}=0.40+Dept/Mag$ [hundred km], $Sig = \Sigma_{m=1,M} \sigma_{H_m}/M$, $\sigma_{H_m} = sqrt(\Sigma_{i=1,N}(H_i-\underline{H}_m)^2)/N$, $\underline{H}_m = \Sigma_{i=1,N} H_i /N$, where $H_i$ is the geomagnetic vector projection, measured 2.4 times per second, M=1440, N=144.

The earthquake precursor (geomagnetic quake) is Sig "irregularity". The vertical error of Mag is distance [Hundred km]. The approximate time window for incoming event is defined from the next Tidal potential (Venedikov et al model) minimum (+/-1day) or maximum (+/-2). EQ data from http://wwwneic.cr.usgs.gov/neis/bulletin/ and GPhI,BAS,Sofia. Reliability time control http://www.emsc-csem.org/.

Figure 20. The relaibilty of earthquake time prediction, Sofia region, 2003

**Acknowledgements** to Boris Vasiliev, JINR, Dubna for magnetometer, Angel Venedikov for Tidal cod VAV, Emil Botev for GPhI, BAS, Sofia almost real time earthquake data, Jeffrey Love for Alaska, College second geomagnetic data, INRNE, BAS, Sofia for support, Cht. S. Mavrodiev, J. Tsatsaragos, A. Damianova and many other colleagues for help, interest, critical and constructive discussions.